\pdfoutput=1
\documentclass[a4paper,11pt]{article}
\usepackage{pos}

\providecommand{\repositoryInformationSetup}{} %
\repositoryInformationSetup

\usepackage{graphicx}
\usepackage[T1]{fontenc}
\usepackage[utf8]{inputenc}
\usepackage[UKenglish]{babel}
\usepackage{placeins}
\usepackage[shortlabels]{enumitem}
\usepackage[binary-units=true]{siunitx}
\usepackage{color}
\usepackage{amsmath}
\usepackage{amsthm}
\usepackage{mathtools}
\usepackage{bbold}
\usepackage{dsfont}
\usepackage{braket}
\usepackage[ruled,lined]{algorithm2e}
\usepackage{nicefrac}
\usepackage{cleveref}
\usepackage[final]{changes}

\graphicspath{{figures/},{inkscape/},{data/}}

\crefformat{footnote}{#2\footnotemark[#1]#3}

\DeclareMathOperator{\diag}{diag}

\newcommand{\del}[2]{\ensuremath{\frac{\partial #1}{\partial#2}}}
\newcommand{\eto}[1]{\ensuremath{\mathrm{e}^{#1}}}
\newcommand{\trans}{\ensuremath{\mathsf{T}}}

\newcommand{\tint}{\ensuremath{\tau_\text{int}}}

\newcommand{\erwartung}[1]{\ensuremath{\left\langle#1\right\rangle}}

\title{The Physicist's Guide to the HMC}

\author*[a]{Johann Ostmeyer}

\affiliation[a]{
	Helmholtz-Institut f\"{u}r Strahlen- und
	Kernphysik,
	University of Bonn, 53115 Bonn, Germany
}
 
\emailAdd{ostmeyer@hiskp.uni-bonn.de}

\abstract{
	The hybrid Monte Carlo (HMC) algorithm is arguably the most efficient sampling method for general probability distributions of continuous variables. Together with exact Fourier acceleration (EFA) the HMC becomes equivalent to direct sampling for quadratic actions $S(x)=\frac12 x^\mathsf{T} M x$ (i.e.\ normal distributions $x\sim \mathrm{e}^{-S(x)}$), only perturbatively worse for perturbative deviations of the action from the quadratic case, and it remains viable for arbitrary actions. In this work the most recent improvements of the HMC including EFA and radial updates are collected into a numerical recipe.
}

\FullConference{The 41st International Symposium on Lattice Field Theory (LATTICE2024)\\
28 July - 3 August 2024\\
Liverpool, UK\\}

\makeatletter%
\makeatletter%
\begin{document}

\maketitle

\section{Introduction}

The hybrid Monte Carlo (HMC) algorithm~\cite{Duane1987} (often Hamiltonian Monte Carlo) allows to sample continuous random variables in high dimensions efficiently. The aim of this work is to explain the HMC in a maximally concise way, focusing on the practical application and referring to existing literature, mainly Ref.~\cite{ostmeyer2024minimal}, for all the derivations. The author hopes to provide a useful guide not only to fellow physicist, but to everyone who wants to use the HMC without having to navigate the galactic amount of related literature.

For this reason, a historical approach is avoided deliberately. Instead, progressively more complicated probability distributions are considered. We start with the (multivariate) normal distribution and add small perturbations to it in \cref{sec:harmonic}. Strong deviations from the normal distribution are pondered in \cref{sec:anharmonic}. Numerical examples are provided to highlight the importance of specific features of the algorithm. Again, the background and technical details of the respective physical problems are omitted. Lattice gauge theories are briefly commented on, but they are not the main focus of this work. 
\section{Harmonically dominated action}\label{sec:harmonic}

Write the probability distribution we want to sample from as $P(x)\propto \eto{-S(x)}$.
Let us start with the case that the action $S(x)$ can be written as
\begin{align}
	S(x) &= \frac12 x^\trans M x + V(x)
\end{align}
and the harmonic part is dominating over small anharmonic perturbations $x^\trans M x\gg |V(x)|$. In this case we want a sampling algorithm that can sample directly from the normal distribution defined by the harmonic part of the action, but that also takes the anharmonic part into account. The HMC algorithm with exact Fourier acceleration EFA~\cite{ostmeyer2024minimal} summarised in \cref{alg:efa-hmc} fulfils these requirements. It guarantees reliable sampling of $P(x)$ with minimal autocorrelation and proceeds as follows.

\begin{algorithm*}[tb]
	\caption{Full HMC trajectory update with EFA (alg.~\ref{alg:efa}) given an integrator (e.g.\ leap-frog, alg.~\ref{alg:leap-frog}).}\label{alg:efa-hmc}
	\SetKwInOut{Input}{input}
	\SetKwInOut{Params}{parameters}
	\SetKwInOut{Output}{output}
	\Input{initial fields $x^\text{i}$, molecular dynamics steps $N_\text{MD}$, trajectory length $T=\frac\pi2$}
	\Params{harmonic matrix $M=\Omega\cdot\diag(\omega^2)\cdot\Omega^\dagger$, anharmonic potential $V$}
	\Output{final fields $x^\text{f}$}
	$x \gets x^\text{i}$\;
	sample $r\sim \mathcal{N}(0,1)^{\dim(M)}$ \tcp*{standard normal distribution}
	$p \gets \Omega\cdot\diag(\omega)\cdot\Omega^\dagger \cdot r$ \tcp*{any realisation of $p\gets\sqrt{M}\cdot r$ can be used}
	$\mathcal{H}^\text{i} \gets \frac12 r^2 + \frac12 x^\trans M x + V(x)$ \tcp*{use $p^\trans M^{-1} p = r^2$}
	\For{$\tau \gets 1\dots N_\text{MD}$}{
		$(x,p) \gets \text{integrator}\left(x,p,\nicefrac{T}{N_\text{MD}}\right)$\;
	}
	$\mathcal{H}^\text{f} \gets \frac12 p^\trans M^{-1} p + \frac12 x^\trans M x + V(x)$\;
	$\Delta \mathcal{H} \gets \mathcal{H}^\text{f}-\mathcal{H}^\text{i}$\;
	\uIf(\tcp*[f]{uniform distribution}){$\eto{-\Delta \mathcal{H}}\ge \mathcal{U}_{[0,1]}$}{$x^\text{f} \gets x$\;}
	\Else{$x^\text{f} \gets x^\text{i}$\;}
\end{algorithm*}

Start from some initial configuration $x$ that is to be updated in a so-called trajectory. Define the Hamiltonian
\begin{align}
	\mathcal{H} &= \frac12 p^\trans M^{-1} p + S(x)\label{eq:opt_hamilton}
\end{align}
and set the trajectory length $T=\frac\pi2$. Now sample the canonical momenta $p\sim \eto{-\frac12 p^\trans M^{-1} p}$ and numerically solve the equations of motion (EOM)
\begin{align}
	\dot x &= \phantom{-}\del{\mathcal{H}}{p}\,,\\
	\dot p &= - \del{\mathcal{H}}{x}
\end{align}
using a symplectic integrator like leap-frog~\cite{OMELYAN2003272,trotter_omelyan}, see alg.~\ref{alg:leap-frog}. The harmonic contributions to the EOM can be solved exactly as described in \cref{alg:efa}. In physics, many problems are translationally invariant and the matrix $M$ can therefore be diagonalised using a Fourier transformation, name-giving for the Fourier acceleration (FA)~\cite{PhysRevD.32.2736}. The resulting configuration is accepted with the probability
\begin{align}
	p_\text{acc} &= \min\left(1, \eto{-\Delta \mathcal{H}}\right)\,,
\end{align}
where $\Delta \mathcal{H}$ denotes the change of the Hamiltonian over the trajectory.

An exact solution of the EOM would preserve the Hamiltonian and thus lead to $\Delta \mathcal{H}=0$ and $p_\text{acc}=1$. Therefore, when the perturbation $V(x)$ is very small, the exact solution of the harmonic part in the EFA allows to achieve very high acceptance rates with a single leap-frog step. With increasing $V(x)$ this approximation becomes less accurate and the acceptance rate drops. If the acceptance drops below ca.\ $70\%$, the number of integration steps should be increased without changing the trajectory length. An acceptance of $65\%$ to $80\%$ (higher for higher order integrators~\cite{Takaishi:1999bi}) is a good compromise between cheap trajectories and low autocorrelation~\cite{Neal:2011mrf}.

\begin{algorithm*}[tb]
	\caption{Single update step with the leap-frog integrator and EFA (alg.~\ref{alg:efa}).}\label{alg:leap-frog}
	\SetKwInOut{Input}{input}
	\SetKwInOut{Params}{parameters}
	\SetKwInOut{Output}{output}
	\Input{initial fields $x^0$, momenta $p^0$, time step $h$}
	\Params{anharmonic forces $-\nabla V$}
	\Output{final fields $x(h)$ and momenta $p(h)$}
	$(x,p) \gets \text{EFA}\left(x^0,p^0,\nicefrac h2\right)$\;
	$p \gets p - h\cdot\nabla V(x)$\;
	$\left(x(h),p(h)\right) \gets \text{EFA}\left(x,p,\nicefrac h2\right)$\;
\end{algorithm*}

\begin{algorithm*}[tb]
	\caption{Single time step using exact Fourier acceleration (EFA).}\label{alg:efa}
	\SetKwInOut{Input}{input}
	\SetKwInOut{Params}{parameters}
	\SetKwInOut{Output}{output}
	\Input{initial fields $x^0$, momenta $p^0$, time step $h$}
	\Params{harmonic matrix $M=\Omega\cdot\diag(\omega^2)\cdot\Omega^\dagger$}
	\Output{final fields $x(h)$ and momenta $p(h)$}
	$y^0 \gets \Omega^\dagger \cdot x^0$ \tcp*{$\Omega$ is often a Fourier trafo, thence the name EFA}
	$q^0 \gets \Omega^\dagger \cdot p^0$\;
	\For{$i \gets 1\dots \dim(M)$}{
		$y_i(h) \gets \cos(h)\, y^0_i + \frac{1}{\omega_i^2} \sin(h)\, q^0_i$\;
		$q_i(h) \gets \cos(h)\, q^0_i - \omega_i^2 \sin(h)\, y^0_i$\;
	}
	$x(h) \gets \Omega \cdot y(h)$\;
	$p(h) \gets \Omega \cdot q(h)$\;
\end{algorithm*}

\subsection{Numerical example}\label{sec:harmonic_example}

The Su-Schrieffer-Heeger (SSH) model provides a realistic description of a multitude of semiconductors like the organic crystal Rubrenes~\cite{Ostmeyer:2023azi}. These materials have a low density of charge carriers that move in an environment of almost free phonons, well approximated by a harmonic oscillator. Therefore simulations of the SSH model in this parameter regime fall exactly into the harmonically dominated class discussed above. That is, the HMC with EFA is perfectly applicable.

\begin{figure*}[t]
	\centering
	\resizebox{0.98\textwidth}{!}{{\large%
\begingroup
  \inputencoding{latin1}%
  \makeatletter
  \providecommand\color[2][]{%
    \GenericError{(gnuplot) \space\space\space\@spaces}{%
      Package color not loaded in conjunction with
      terminal option `colourtext'%
    }{See the gnuplot documentation for explanation.%
    }{Either use 'blacktext' in gnuplot or load the package
      color.sty in LaTeX.}%
    \renewcommand\color[2][]{}%
  }%
  \providecommand\includegraphics[2][]{%
    \GenericError{(gnuplot) \space\space\space\@spaces}{%
      Package graphicx or graphics not loaded%
    }{See the gnuplot documentation for explanation.%
    }{The gnuplot epslatex terminal needs graphicx.sty or graphics.sty.}%
    \renewcommand\includegraphics[2][]{}%
  }%
  \providecommand\rotatebox[2]{#2}%
  \@ifundefined{ifGPcolor}{%
    \newif\ifGPcolor
    \GPcolortrue
  }{}%
  \@ifundefined{ifGPblacktext}{%
    \newif\ifGPblacktext
    \GPblacktexttrue
  }{}%
  \let\gplgaddtomacro\g@addto@macro
  \gdef\gplbacktext{}%
  \gdef\gplfronttext{}%
  \makeatother
  \ifGPblacktext
    \def\colorrgb#1{}%
    \def\colorgray#1{}%
  \else
    \ifGPcolor
      \def\colorrgb#1{\color[rgb]{#1}}%
      \def\colorgray#1{\color[gray]{#1}}%
      \expandafter\def\csname LTw\endcsname{\color{white}}%
      \expandafter\def\csname LTb\endcsname{\color{black}}%
      \expandafter\def\csname LTa\endcsname{\color{black}}%
      \expandafter\def\csname LT0\endcsname{\color[rgb]{1,0,0}}%
      \expandafter\def\csname LT1\endcsname{\color[rgb]{0,1,0}}%
      \expandafter\def\csname LT2\endcsname{\color[rgb]{0,0,1}}%
      \expandafter\def\csname LT3\endcsname{\color[rgb]{1,0,1}}%
      \expandafter\def\csname LT4\endcsname{\color[rgb]{0,1,1}}%
      \expandafter\def\csname LT5\endcsname{\color[rgb]{1,1,0}}%
      \expandafter\def\csname LT6\endcsname{\color[rgb]{0,0,0}}%
      \expandafter\def\csname LT7\endcsname{\color[rgb]{1,0.3,0}}%
      \expandafter\def\csname LT8\endcsname{\color[rgb]{0.5,0.5,0.5}}%
    \else
      \def\colorrgb#1{\color{black}}%
      \def\colorgray#1{\color[gray]{#1}}%
      \expandafter\def\csname LTw\endcsname{\color{white}}%
      \expandafter\def\csname LTb\endcsname{\color{black}}%
      \expandafter\def\csname LTa\endcsname{\color{black}}%
      \expandafter\def\csname LT0\endcsname{\color{black}}%
      \expandafter\def\csname LT1\endcsname{\color{black}}%
      \expandafter\def\csname LT2\endcsname{\color{black}}%
      \expandafter\def\csname LT3\endcsname{\color{black}}%
      \expandafter\def\csname LT4\endcsname{\color{black}}%
      \expandafter\def\csname LT5\endcsname{\color{black}}%
      \expandafter\def\csname LT6\endcsname{\color{black}}%
      \expandafter\def\csname LT7\endcsname{\color{black}}%
      \expandafter\def\csname LT8\endcsname{\color{black}}%
    \fi
  \fi
    \setlength{\unitlength}{0.0500bp}%
    \ifx\gptboxheight\undefined%
      \newlength{\gptboxheight}%
      \newlength{\gptboxwidth}%
      \newsavebox{\gptboxtext}%
    \fi%
    \setlength{\fboxrule}{0.5pt}%
    \setlength{\fboxsep}{1pt}%
\begin{picture}(7200.00,5040.00)%
    \gplgaddtomacro\gplbacktext{%
      \csname LTb\endcsname%
      \put(1078,1076){\makebox(0,0)[r]{\strut{}$1$}}%
      \csname LTb\endcsname%
      \put(1078,2012){\makebox(0,0)[r]{\strut{}$10$}}%
      \csname LTb\endcsname%
      \put(1078,2948){\makebox(0,0)[r]{\strut{}$100$}}%
      \csname LTb\endcsname%
      \put(1078,3883){\makebox(0,0)[r]{\strut{}$1000$}}%
      \csname LTb\endcsname%
      \put(1078,4819){\makebox(0,0)[r]{\strut{}$10000$}}%
      \csname LTb\endcsname%
      \put(1210,484){\makebox(0,0){\strut{}$1$}}%
      \csname LTb\endcsname%
      \put(3074,484){\makebox(0,0){\strut{}$10$}}%
      \csname LTb\endcsname%
      \put(4939,484){\makebox(0,0){\strut{}$100$}}%
      \csname LTb\endcsname%
      \put(6803,484){\makebox(0,0){\strut{}$1000$}}%
    }%
    \gplgaddtomacro\gplfronttext{%
      \csname LTb\endcsname%
      \put(209,2761){\rotatebox{-270}{\makebox(0,0){\strut{}$\tau_\text{int}$}}}%
      \put(4006,154){\makebox(0,0){\strut{}$\omega_0 [\si{\milli\eV}]$}}%
      \csname LTb\endcsname%
      \put(2926,4591){\makebox(0,0)[r]{\strut{}EFA, $\erwartung{n_\text{el}}$}}%
      \csname LTb\endcsname%
      \put(2926,4261){\makebox(0,0)[r]{\strut{}no FA, $\erwartung{n_\text{el}}$}}%
      \csname LTb\endcsname%
      \put(2926,3931){\makebox(0,0)[r]{\strut{}prediction}}%
      \csname LTb\endcsname%
      \put(2926,3601){\makebox(0,0)[r]{\strut{}no FA, $\erwartung{n_\text{ph}}$}}%
    }%
    \gplbacktext
    \put(0,0){\includegraphics{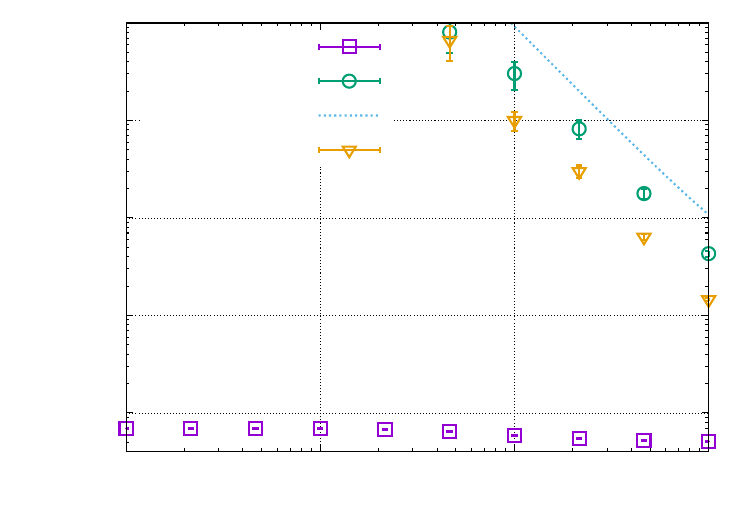}}%
    \gplfronttext
  \end{picture}%
\endgroup
\begingroup
  \inputencoding{latin1}%
  \makeatletter
  \providecommand\color[2][]{%
    \GenericError{(gnuplot) \space\space\space\@spaces}{%
      Package color not loaded in conjunction with
      terminal option `colourtext'%
    }{See the gnuplot documentation for explanation.%
    }{Either use 'blacktext' in gnuplot or load the package
      color.sty in LaTeX.}%
    \renewcommand\color[2][]{}%
  }%
  \providecommand\includegraphics[2][]{%
    \GenericError{(gnuplot) \space\space\space\@spaces}{%
      Package graphicx or graphics not loaded%
    }{See the gnuplot documentation for explanation.%
    }{The gnuplot epslatex terminal needs graphicx.sty or graphics.sty.}%
    \renewcommand\includegraphics[2][]{}%
  }%
  \providecommand\rotatebox[2]{#2}%
  \@ifundefined{ifGPcolor}{%
    \newif\ifGPcolor
    \GPcolortrue
  }{}%
  \@ifundefined{ifGPblacktext}{%
    \newif\ifGPblacktext
    \GPblacktexttrue
  }{}%
  \let\gplgaddtomacro\g@addto@macro
  \gdef\gplbacktext{}%
  \gdef\gplfronttext{}%
  \makeatother
  \ifGPblacktext
    \def\colorrgb#1{}%
    \def\colorgray#1{}%
  \else
    \ifGPcolor
      \def\colorrgb#1{\color[rgb]{#1}}%
      \def\colorgray#1{\color[gray]{#1}}%
      \expandafter\def\csname LTw\endcsname{\color{white}}%
      \expandafter\def\csname LTb\endcsname{\color{black}}%
      \expandafter\def\csname LTa\endcsname{\color{black}}%
      \expandafter\def\csname LT0\endcsname{\color[rgb]{1,0,0}}%
      \expandafter\def\csname LT1\endcsname{\color[rgb]{0,1,0}}%
      \expandafter\def\csname LT2\endcsname{\color[rgb]{0,0,1}}%
      \expandafter\def\csname LT3\endcsname{\color[rgb]{1,0,1}}%
      \expandafter\def\csname LT4\endcsname{\color[rgb]{0,1,1}}%
      \expandafter\def\csname LT5\endcsname{\color[rgb]{1,1,0}}%
      \expandafter\def\csname LT6\endcsname{\color[rgb]{0,0,0}}%
      \expandafter\def\csname LT7\endcsname{\color[rgb]{1,0.3,0}}%
      \expandafter\def\csname LT8\endcsname{\color[rgb]{0.5,0.5,0.5}}%
    \else
      \def\colorrgb#1{\color{black}}%
      \def\colorgray#1{\color[gray]{#1}}%
      \expandafter\def\csname LTw\endcsname{\color{white}}%
      \expandafter\def\csname LTb\endcsname{\color{black}}%
      \expandafter\def\csname LTa\endcsname{\color{black}}%
      \expandafter\def\csname LT0\endcsname{\color{black}}%
      \expandafter\def\csname LT1\endcsname{\color{black}}%
      \expandafter\def\csname LT2\endcsname{\color{black}}%
      \expandafter\def\csname LT3\endcsname{\color{black}}%
      \expandafter\def\csname LT4\endcsname{\color{black}}%
      \expandafter\def\csname LT5\endcsname{\color{black}}%
      \expandafter\def\csname LT6\endcsname{\color{black}}%
      \expandafter\def\csname LT7\endcsname{\color{black}}%
      \expandafter\def\csname LT8\endcsname{\color{black}}%
    \fi
  \fi
    \setlength{\unitlength}{0.0500bp}%
    \ifx\gptboxheight\undefined%
      \newlength{\gptboxheight}%
      \newlength{\gptboxwidth}%
      \newsavebox{\gptboxtext}%
    \fi%
    \setlength{\fboxrule}{0.5pt}%
    \setlength{\fboxsep}{1pt}%
\begin{picture}(7200.00,5040.00)%
    \gplgaddtomacro\gplbacktext{%
      \csname LTb\endcsname%
      \put(1078,1076){\makebox(0,0)[r]{\strut{}$1$}}%
      \csname LTb\endcsname%
      \put(1078,2012){\makebox(0,0)[r]{\strut{}$10$}}%
      \csname LTb\endcsname%
      \put(1078,2948){\makebox(0,0)[r]{\strut{}$100$}}%
      \csname LTb\endcsname%
      \put(1078,3883){\makebox(0,0)[r]{\strut{}$1000$}}%
      \csname LTb\endcsname%
      \put(1078,4819){\makebox(0,0)[r]{\strut{}$10000$}}%
      \csname LTb\endcsname%
      \put(1271,484){\makebox(0,0){\strut{}$0.001$}}%
      \csname LTb\endcsname%
      \put(5466,484){\makebox(0,0){\strut{}$0.01$}}%
    }%
    \gplgaddtomacro\gplfronttext{%
      \csname LTb\endcsname%
      \put(209,2761){\rotatebox{-270}{\makebox(0,0){\strut{}$\tau_\text{int}$}}}%
      \put(4006,154){\makebox(0,0){\strut{}$1/N_t$}}%
      \csname LTb\endcsname%
      \put(2926,3256){\makebox(0,0)[r]{\strut{}EFA, $\erwartung{n_\text{el}}$}}%
      \csname LTb\endcsname%
      \put(2926,2926){\makebox(0,0)[r]{\strut{}no FA, $\erwartung{n_\text{el}}$}}%
      \csname LTb\endcsname%
      \put(2926,2596){\makebox(0,0)[r]{\strut{}prediction}}%
      \csname LTb\endcsname%
      \put(2926,2266){\makebox(0,0)[r]{\strut{}no FA, $\erwartung{n_\text{ph}}$}}%
    }%
    \gplbacktext
    \put(0,0){\includegraphics{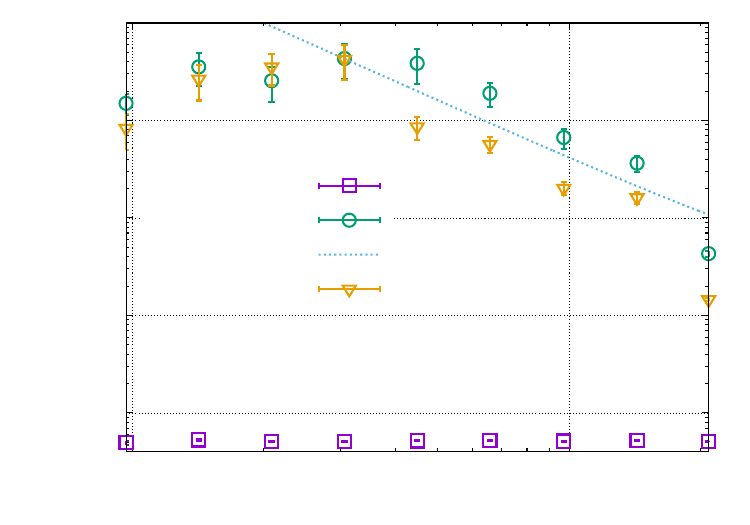}}%
    \gplfronttext
  \end{picture}%
\endgroup
}}
	\caption{Integrated autocorrelation time $\tau_\text{int}$ of the phonon $\erwartung{n_\text{ph}}$ and electron $\erwartung{n_\text{el}}$ number expectation values in the 2+1D SSH model using HMC simulations~\cite{Ostmeyer:2023azi} on a $10\times 10$ lattice with $N_t$ imaginary time slices. With EFA $\tint$ for both observables coincides. The physical parameters are realistic for the organic semiconductor Rubrene at room temperature $\beta=\SI{40}{\per\eV}$ (see tab.~I of the supl.\ mat.\ in~\cite{Ostmeyer:2023azi} with $\mu=-2\sum_\alpha|J_\alpha|$). Left: different free phonon frequencies $\omega_0$ at constant $N_t=48$ ($\omega_0\approx \SI{6}{\milli\eV}$ is the physical value); Right: different $N_t$ at fixed $\omega_0=\SI{1}{\eV}$. All simulations have similar acceptance ($\gtrsim 80\%$) and compute time per measurement. The dashed line shows the proportionality prediction for $\tau_\text{int}$ from corollary 1 in Ref.~\cite{ostmeyer2024minimal}.}
	\label{fig:ssh_rubrene}
\end{figure*}

Figure~\ref{fig:ssh_rubrene} demonstrates the efficiency of the HMC with EFA (alg.~\ref{alg:efa-hmc}). The integrated autocorrelation time $\tint\ge\num{0.5}$ is shown as a figure of merit. It quantifies how many trajectories are required to obtain a statistically independent measurement~\cite{WOLFF2004143}. Since EFA allows almost uncorrelated sampling, it reliably leads to minimal autocorrelations $\tint \lesssim\num{0.7}$, even when the continuum limit is approached. These simulations can be compared to the ``classical'' HMC algorithm without FA, i.e.\ using $\mathcal{H}=\frac12 p^2 +S(x)$ instead of the optimal choice from equation~\eqref{eq:opt_hamilton}. It becomes clear immediately that simulations without FA lead to prohibitively long autocorrelation times, diverging towards the continuum limit.

\section{Generic action}\label{sec:anharmonic}

A major advantage of the HMC over most competitor algorithms is its applicability to any sampling problem involving continuous variables. That said, there are several essential adjustments that have to be introduced when the action is not harmonically dominated.

\subsection{Necessary adjustments}

\subsubsection{Randomise the trajectory length}

Once the eigenmodes of the HMC dynamics are not known, it is theoretically possible to choose a trajectory length $T\approx0\pmod{2\pi}$ for one of the modes. This mode would then decorrelate very slowly. To avoid this, the trajectory length should be randomised~\cite{Mackenzie:1989us,apers2022hamiltonian}. A realisation that also takes care of maintaining a good trajectory length on average is the No-U-turn sampling~\cite{NUTS-2014}, though typically much simpler schemes yield comparable results.

\subsubsection{Use long trajectories}

\begin{figure*}[t]
	\centering
	\resizebox{0.98\textwidth}{!}{{\large%
\begingroup
  \inputencoding{latin1}%
  \makeatletter
  \providecommand\color[2][]{%
    \GenericError{(gnuplot) \space\space\space\@spaces}{%
      Package color not loaded in conjunction with
      terminal option `colourtext'%
    }{See the gnuplot documentation for explanation.%
    }{Either use 'blacktext' in gnuplot or load the package
      color.sty in LaTeX.}%
    \renewcommand\color[2][]{}%
  }%
  \providecommand\includegraphics[2][]{%
    \GenericError{(gnuplot) \space\space\space\@spaces}{%
      Package graphicx or graphics not loaded%
    }{See the gnuplot documentation for explanation.%
    }{The gnuplot epslatex terminal needs graphicx.sty or graphics.sty.}%
    \renewcommand\includegraphics[2][]{}%
  }%
  \providecommand\rotatebox[2]{#2}%
  \@ifundefined{ifGPcolor}{%
    \newif\ifGPcolor
    \GPcolortrue
  }{}%
  \@ifundefined{ifGPblacktext}{%
    \newif\ifGPblacktext
    \GPblacktexttrue
  }{}%
  \let\gplgaddtomacro\g@addto@macro
  \gdef\gplbacktext{}%
  \gdef\gplfronttext{}%
  \makeatother
  \ifGPblacktext
    \def\colorrgb#1{}%
    \def\colorgray#1{}%
  \else
    \ifGPcolor
      \def\colorrgb#1{\color[rgb]{#1}}%
      \def\colorgray#1{\color[gray]{#1}}%
      \expandafter\def\csname LTw\endcsname{\color{white}}%
      \expandafter\def\csname LTb\endcsname{\color{black}}%
      \expandafter\def\csname LTa\endcsname{\color{black}}%
      \expandafter\def\csname LT0\endcsname{\color[rgb]{1,0,0}}%
      \expandafter\def\csname LT1\endcsname{\color[rgb]{0,1,0}}%
      \expandafter\def\csname LT2\endcsname{\color[rgb]{0,0,1}}%
      \expandafter\def\csname LT3\endcsname{\color[rgb]{1,0,1}}%
      \expandafter\def\csname LT4\endcsname{\color[rgb]{0,1,1}}%
      \expandafter\def\csname LT5\endcsname{\color[rgb]{1,1,0}}%
      \expandafter\def\csname LT6\endcsname{\color[rgb]{0,0,0}}%
      \expandafter\def\csname LT7\endcsname{\color[rgb]{1,0.3,0}}%
      \expandafter\def\csname LT8\endcsname{\color[rgb]{0.5,0.5,0.5}}%
    \else
      \def\colorrgb#1{\color{black}}%
      \def\colorgray#1{\color[gray]{#1}}%
      \expandafter\def\csname LTw\endcsname{\color{white}}%
      \expandafter\def\csname LTb\endcsname{\color{black}}%
      \expandafter\def\csname LTa\endcsname{\color{black}}%
      \expandafter\def\csname LT0\endcsname{\color{black}}%
      \expandafter\def\csname LT1\endcsname{\color{black}}%
      \expandafter\def\csname LT2\endcsname{\color{black}}%
      \expandafter\def\csname LT3\endcsname{\color{black}}%
      \expandafter\def\csname LT4\endcsname{\color{black}}%
      \expandafter\def\csname LT5\endcsname{\color{black}}%
      \expandafter\def\csname LT6\endcsname{\color{black}}%
      \expandafter\def\csname LT7\endcsname{\color{black}}%
      \expandafter\def\csname LT8\endcsname{\color{black}}%
    \fi
  \fi
    \setlength{\unitlength}{0.0500bp}%
    \ifx\gptboxheight\undefined%
      \newlength{\gptboxheight}%
      \newlength{\gptboxwidth}%
      \newsavebox{\gptboxtext}%
    \fi%
    \setlength{\fboxrule}{0.5pt}%
    \setlength{\fboxsep}{1pt}%
\begin{picture}(7200.00,5040.00)%
    \gplgaddtomacro\gplbacktext{%
      \csname LTb\endcsname%
      \put(814,1117){\makebox(0,0)[r]{\strut{}$1$}}%
      \csname LTb\endcsname%
      \put(814,2489){\makebox(0,0)[r]{\strut{}$10$}}%
      \csname LTb\endcsname%
      \put(814,3860){\makebox(0,0)[r]{\strut{}$100$}}%
      \csname LTb\endcsname%
      \put(946,484){\makebox(0,0){\strut{}$0$}}%
      \csname LTb\endcsname%
      \put(1678,484){\makebox(0,0){\strut{}$0.1$}}%
      \csname LTb\endcsname%
      \put(2410,484){\makebox(0,0){\strut{}$0.2$}}%
      \csname LTb\endcsname%
      \put(3142,484){\makebox(0,0){\strut{}$0.3$}}%
      \csname LTb\endcsname%
      \put(3875,484){\makebox(0,0){\strut{}$0.4$}}%
      \csname LTb\endcsname%
      \put(4607,484){\makebox(0,0){\strut{}$0.5$}}%
      \csname LTb\endcsname%
      \put(5339,484){\makebox(0,0){\strut{}$0.6$}}%
      \csname LTb\endcsname%
      \put(6071,484){\makebox(0,0){\strut{}$0.7$}}%
      \csname LTb\endcsname%
      \put(6803,484){\makebox(0,0){\strut{}$0.8$}}%
    }%
    \gplgaddtomacro\gplfronttext{%
      \csname LTb\endcsname%
      \put(209,2761){\rotatebox{-270}{\makebox(0,0){\strut{}$\tau_\text{int}$}}}%
      \put(3874,154){\makebox(0,0){\strut{}$J$}}%
      \csname LTb\endcsname%
      \put(5816,1592){\makebox(0,0)[r]{\strut{}EFA, $T=\nicefrac\pi6$}}%
      \csname LTb\endcsname%
      \put(5816,1262){\makebox(0,0)[r]{\strut{}EFA, $T=\nicefrac\pi2$}}%
      \csname LTb\endcsname%
      \put(5816,932){\makebox(0,0)[r]{\strut{}prediction for $T=\nicefrac\pi6$}}%
    }%
    \gplbacktext
    \put(0,0){\includegraphics{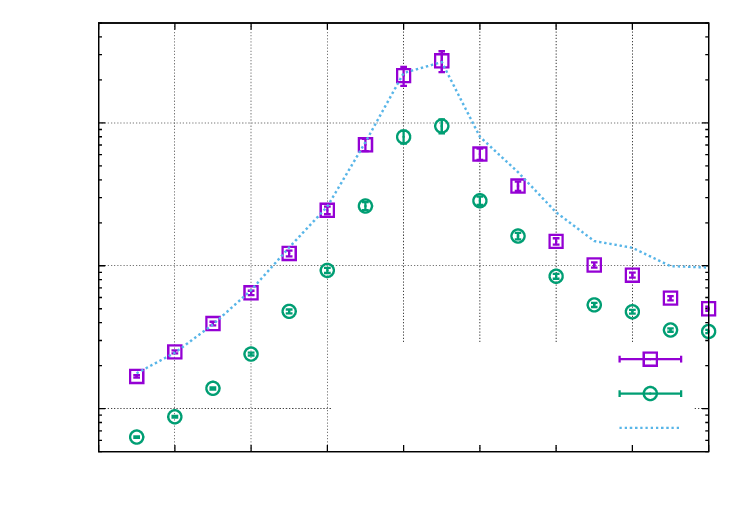}}%
    \gplfronttext
  \end{picture}%
\endgroup
\begingroup
  \inputencoding{latin1}%
  \makeatletter
  \providecommand\color[2][]{%
    \GenericError{(gnuplot) \space\space\space\@spaces}{%
      Package color not loaded in conjunction with
      terminal option `colourtext'%
    }{See the gnuplot documentation for explanation.%
    }{Either use 'blacktext' in gnuplot or load the package
      color.sty in LaTeX.}%
    \renewcommand\color[2][]{}%
  }%
  \providecommand\includegraphics[2][]{%
    \GenericError{(gnuplot) \space\space\space\@spaces}{%
      Package graphicx or graphics not loaded%
    }{See the gnuplot documentation for explanation.%
    }{The gnuplot epslatex terminal needs graphicx.sty or graphics.sty.}%
    \renewcommand\includegraphics[2][]{}%
  }%
  \providecommand\rotatebox[2]{#2}%
  \@ifundefined{ifGPcolor}{%
    \newif\ifGPcolor
    \GPcolortrue
  }{}%
  \@ifundefined{ifGPblacktext}{%
    \newif\ifGPblacktext
    \GPblacktexttrue
  }{}%
  \let\gplgaddtomacro\g@addto@macro
  \gdef\gplbacktext{}%
  \gdef\gplfronttext{}%
  \makeatother
  \ifGPblacktext
    \def\colorrgb#1{}%
    \def\colorgray#1{}%
  \else
    \ifGPcolor
      \def\colorrgb#1{\color[rgb]{#1}}%
      \def\colorgray#1{\color[gray]{#1}}%
      \expandafter\def\csname LTw\endcsname{\color{white}}%
      \expandafter\def\csname LTb\endcsname{\color{black}}%
      \expandafter\def\csname LTa\endcsname{\color{black}}%
      \expandafter\def\csname LT0\endcsname{\color[rgb]{1,0,0}}%
      \expandafter\def\csname LT1\endcsname{\color[rgb]{0,1,0}}%
      \expandafter\def\csname LT2\endcsname{\color[rgb]{0,0,1}}%
      \expandafter\def\csname LT3\endcsname{\color[rgb]{1,0,1}}%
      \expandafter\def\csname LT4\endcsname{\color[rgb]{0,1,1}}%
      \expandafter\def\csname LT5\endcsname{\color[rgb]{1,1,0}}%
      \expandafter\def\csname LT6\endcsname{\color[rgb]{0,0,0}}%
      \expandafter\def\csname LT7\endcsname{\color[rgb]{1,0.3,0}}%
      \expandafter\def\csname LT8\endcsname{\color[rgb]{0.5,0.5,0.5}}%
    \else
      \def\colorrgb#1{\color{black}}%
      \def\colorgray#1{\color[gray]{#1}}%
      \expandafter\def\csname LTw\endcsname{\color{white}}%
      \expandafter\def\csname LTb\endcsname{\color{black}}%
      \expandafter\def\csname LTa\endcsname{\color{black}}%
      \expandafter\def\csname LT0\endcsname{\color{black}}%
      \expandafter\def\csname LT1\endcsname{\color{black}}%
      \expandafter\def\csname LT2\endcsname{\color{black}}%
      \expandafter\def\csname LT3\endcsname{\color{black}}%
      \expandafter\def\csname LT4\endcsname{\color{black}}%
      \expandafter\def\csname LT5\endcsname{\color{black}}%
      \expandafter\def\csname LT6\endcsname{\color{black}}%
      \expandafter\def\csname LT7\endcsname{\color{black}}%
      \expandafter\def\csname LT8\endcsname{\color{black}}%
    \fi
  \fi
    \setlength{\unitlength}{0.0500bp}%
    \ifx\gptboxheight\undefined%
      \newlength{\gptboxheight}%
      \newlength{\gptboxwidth}%
      \newsavebox{\gptboxtext}%
    \fi%
    \setlength{\fboxrule}{0.5pt}%
    \setlength{\fboxsep}{1pt}%
\begin{picture}(7200.00,5040.00)%
    \gplgaddtomacro\gplbacktext{%
      \csname LTb\endcsname%
      \put(682,704){\makebox(0,0)[r]{\strut{}$0$}}%
      \csname LTb\endcsname%
      \put(682,1527){\makebox(0,0)[r]{\strut{}$5$}}%
      \csname LTb\endcsname%
      \put(682,2350){\makebox(0,0)[r]{\strut{}$10$}}%
      \csname LTb\endcsname%
      \put(682,3173){\makebox(0,0)[r]{\strut{}$15$}}%
      \csname LTb\endcsname%
      \put(682,3996){\makebox(0,0)[r]{\strut{}$20$}}%
      \csname LTb\endcsname%
      \put(682,4819){\makebox(0,0)[r]{\strut{}$25$}}%
      \csname LTb\endcsname%
      \put(814,484){\makebox(0,0){\strut{}$0$}}%
      \csname LTb\endcsname%
      \put(1563,484){\makebox(0,0){\strut{}$0.2$}}%
      \csname LTb\endcsname%
      \put(2311,484){\makebox(0,0){\strut{}$0.4$}}%
      \csname LTb\endcsname%
      \put(3060,484){\makebox(0,0){\strut{}$0.6$}}%
      \csname LTb\endcsname%
      \put(3809,484){\makebox(0,0){\strut{}$0.8$}}%
      \csname LTb\endcsname%
      \put(4557,484){\makebox(0,0){\strut{}$1$}}%
      \csname LTb\endcsname%
      \put(5306,484){\makebox(0,0){\strut{}$1.2$}}%
      \csname LTb\endcsname%
      \put(6054,484){\makebox(0,0){\strut{}$1.4$}}%
      \csname LTb\endcsname%
      \put(6803,484){\makebox(0,0){\strut{}$1.6$}}%
    }%
    \gplgaddtomacro\gplfronttext{%
      \csname LTb\endcsname%
      \put(209,2761){\rotatebox{-270}{\makebox(0,0){\strut{}$\tau_\text{int}$}}}%
      \put(3808,154){\makebox(0,0){\strut{}$T$}}%
      \csname LTb\endcsname%
      \put(5816,4591){\makebox(0,0)[r]{\strut{}EFA, $J=\num{0.2}$}}%
      \csname LTb\endcsname%
      \put(5816,4261){\makebox(0,0)[r]{\strut{}prediction}}%
    }%
    \gplbacktext
    \put(0,0){\includegraphics{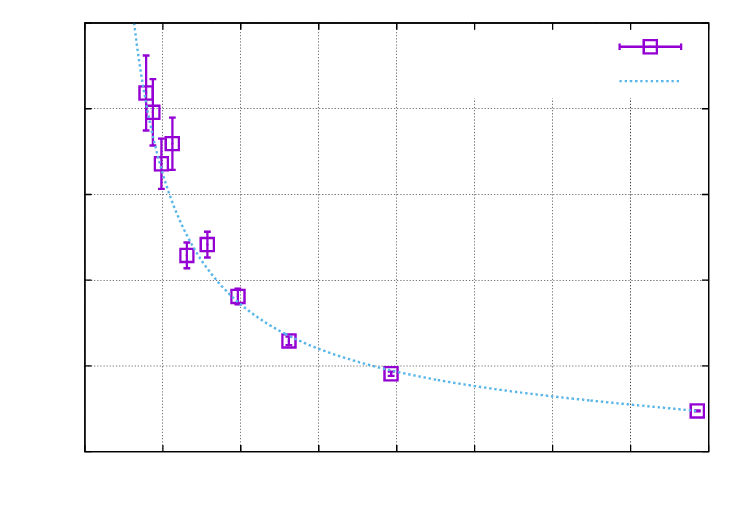}}%
    \gplfronttext
  \end{picture}%
\endgroup
}}
	\caption{Integrated autocorrelation time $\tau_\text{int}$ of the absolute value of the magnetisation $|m|$ in the 2D Ising model using HMC simulations~\cite{ising} on a $15\times 15$ lattice. Left: different coupling strengths ($J\approx \num{0.44}$ is the critical coupling~\cite{onsager_2d_solution}); Right: weak coupling $J=\num{0.2}$ and different trajectory lengths $T$. All simulations used EFA and have similar acceptance $\gtrsim 80\%$. The measurement frequency has been adjusted so that the HMC time between measurements is always the same. The dashed line shows the prediction for $\tau_\text{int}(T)=\tau_\text{int}(T=\nicefrac{\pi}{2})\cdot \left(1-\cos(T)^{\nicefrac{\pi}{2T}}\right)^{-1}$ from corollary 2 in Ref.~\cite{ostmeyer2024minimal}.}
	\label{fig:ising_var_J}
\end{figure*}

Crucially, the (average) trajectory length has to be chosen sufficiently long even if the action is not harmonically dominated~\cite{apers2022hamiltonian}. Here, sufficiently long means that the autocorrelation time cannot be reduced significantly by further increasing the trajectory length. Short trajectories result in a highly inefficient diffusive regime.
The importance of an appropriate trajectory length is demonstrated in figure~\ref{fig:ising_var_J}. Over the entire parameter range of the simulated Ising model spanning from almost harmonic for couplings $J\lesssim\num{0.1}$ to strongly anharmonic for $J\gtrsim\num{0.5}$ we find that too short trajectories increase the integrated autocorrelation time $\tint$ and thus reduce sampling efficiency. 

At the same time, figure~\ref{fig:ising_var_J} vividly demonstrates a limitation of the HMC, be it with EFA or without. HMC simulations of the Ising model experience critical slowing down, that is the autocorrelation time diverges with the volume at the point of the phase transition $J\approx\num{0.44}$. This is caused by potential barriers in the phase space that cannot be overcome efficiently by a continuous evolution.

\subsubsection{Include radial updates on non-compact spaces}

On non-compact spaces (including all cases discussed so far) the HMC is not guaranteed to converge exponentially quickly to the desired probability distribution. A simple fix to this problem is the introduction of so-called radial updates at regular intervals~\cite{original_radial_update,Ostmeyer:2024gnh}, e.g.\ after every HMC trajectory. In some cases radial updates can also restore ergodicity to simulations involving actions with potential barriers~\cite{Temmen:2024pcm,Temmen_ergodicity}. For polynomial actions the radial update is summarised in algorithm~\ref{alg:poly_pot}. It amounts to a multiplicative update of the current state followed by an accept/reject step. An efficient generalisation for arbitrary actions is provided in Ref.~\cite{Ostmeyer:2024gnh}.

\begin{algorithm*}[tb]
	\caption{Radial update from Ref.~\cite{Ostmeyer:2024gnh} sampling $x\in \mathds{R}^d$ from the probability distribution $p(x)\propto\eto{-S(x)}$ generated by a polynomial action $S$.}\label{alg:poly_pot}
	\SetKwInOut{Input}{input}
	\SetKwInOut{Params}{parameters}
	\SetKwInOut{Output}{output}
	\Params{dimension $d$, action $S$ with $S(x)\approx c|x|^a$ for large $|x|$}
	\Input{initial vector $x^\text{i}\in X$, standard deviation $\sigma$ (default $\sigma = \sqrt{\frac{2}{ad}}$)}
	\Output{final vector $x^\text{f}\in X$}
	sample $\gamma\sim \mathcal{N}(0,\sigma^2)$ \tcp*{normal distribution}
	$x \gets x^\text{i}\cdot \eto{\gamma}$\;
	$\Delta S \gets S(x)-S(x^\text{i})$\;
	\uIf(\tcp*[f]{uniform distribution}){$\eto{-\Delta S+d\gamma}\ge \mathcal{U}_{[0,1]}$}{$x^\text{f} \gets x$\;}
	\Else{$x^\text{f} \gets x^\text{i}$\;}
\end{algorithm*}

\subsubsection{Regularise the kinetic term}

In the strongly anharmonic case $|V(x)|\gg |x^\trans M x|$ there is no guarantee for $M$ to be positive definite, much less a good approximation of the action. Therefore the first term of the Hamiltonian in equation~\eqref{eq:opt_hamilton} (the `kinetic energy') might not be well-defined or simply a bad choice.
Regularisations that restore correctness to the algorithm are readily available. For instance the Hamiltonian
\begin{align}
	\mathcal{H} &= \frac12 p^\trans \left(M + \mu\right)^{-1}p + S(x)
\end{align}
is a valid choice for any regulator $\mu\ge0$ large enough so that $M+\mu$ is strictly positive definite.
Unfortunately, it is impossible to derive the optimal kinetic term without further knowledge of the anharmonic potential $V(x)$.

\subsection{Lattice gauge theory}

\begin{figure*}[t]
	\centering
	\resizebox{0.98\textwidth}{!}{{\large%
\begingroup
  \inputencoding{latin1}%
  \makeatletter
  \providecommand\color[2][]{%
    \GenericError{(gnuplot) \space\space\space\@spaces}{%
      Package color not loaded in conjunction with
      terminal option `colourtext'%
    }{See the gnuplot documentation for explanation.%
    }{Either use 'blacktext' in gnuplot or load the package
      color.sty in LaTeX.}%
    \renewcommand\color[2][]{}%
  }%
  \providecommand\includegraphics[2][]{%
    \GenericError{(gnuplot) \space\space\space\@spaces}{%
      Package graphicx or graphics not loaded%
    }{See the gnuplot documentation for explanation.%
    }{The gnuplot epslatex terminal needs graphicx.sty or graphics.sty.}%
    \renewcommand\includegraphics[2][]{}%
  }%
  \providecommand\rotatebox[2]{#2}%
  \@ifundefined{ifGPcolor}{%
    \newif\ifGPcolor
    \GPcolortrue
  }{}%
  \@ifundefined{ifGPblacktext}{%
    \newif\ifGPblacktext
    \GPblacktexttrue
  }{}%
  \let\gplgaddtomacro\g@addto@macro
  \gdef\gplbacktext{}%
  \gdef\gplfronttext{}%
  \makeatother
  \ifGPblacktext
    \def\colorrgb#1{}%
    \def\colorgray#1{}%
  \else
    \ifGPcolor
      \def\colorrgb#1{\color[rgb]{#1}}%
      \def\colorgray#1{\color[gray]{#1}}%
      \expandafter\def\csname LTw\endcsname{\color{white}}%
      \expandafter\def\csname LTb\endcsname{\color{black}}%
      \expandafter\def\csname LTa\endcsname{\color{black}}%
      \expandafter\def\csname LT0\endcsname{\color[rgb]{1,0,0}}%
      \expandafter\def\csname LT1\endcsname{\color[rgb]{0,1,0}}%
      \expandafter\def\csname LT2\endcsname{\color[rgb]{0,0,1}}%
      \expandafter\def\csname LT3\endcsname{\color[rgb]{1,0,1}}%
      \expandafter\def\csname LT4\endcsname{\color[rgb]{0,1,1}}%
      \expandafter\def\csname LT5\endcsname{\color[rgb]{1,1,0}}%
      \expandafter\def\csname LT6\endcsname{\color[rgb]{0,0,0}}%
      \expandafter\def\csname LT7\endcsname{\color[rgb]{1,0.3,0}}%
      \expandafter\def\csname LT8\endcsname{\color[rgb]{0.5,0.5,0.5}}%
    \else
      \def\colorrgb#1{\color{black}}%
      \def\colorgray#1{\color[gray]{#1}}%
      \expandafter\def\csname LTw\endcsname{\color{white}}%
      \expandafter\def\csname LTb\endcsname{\color{black}}%
      \expandafter\def\csname LTa\endcsname{\color{black}}%
      \expandafter\def\csname LT0\endcsname{\color{black}}%
      \expandafter\def\csname LT1\endcsname{\color{black}}%
      \expandafter\def\csname LT2\endcsname{\color{black}}%
      \expandafter\def\csname LT3\endcsname{\color{black}}%
      \expandafter\def\csname LT4\endcsname{\color{black}}%
      \expandafter\def\csname LT5\endcsname{\color{black}}%
      \expandafter\def\csname LT6\endcsname{\color{black}}%
      \expandafter\def\csname LT7\endcsname{\color{black}}%
      \expandafter\def\csname LT8\endcsname{\color{black}}%
    \fi
  \fi
    \setlength{\unitlength}{0.0500bp}%
    \ifx\gptboxheight\undefined%
      \newlength{\gptboxheight}%
      \newlength{\gptboxwidth}%
      \newsavebox{\gptboxtext}%
    \fi%
    \setlength{\fboxrule}{0.5pt}%
    \setlength{\fboxsep}{1pt}%
\begin{picture}(7200.00,5040.00)%
    \gplgaddtomacro\gplbacktext{%
      \csname LTb\endcsname%
      \put(814,1242){\makebox(0,0)[r]{\strut{}$1$}}%
      \csname LTb\endcsname%
      \put(814,3031){\makebox(0,0)[r]{\strut{}$10$}}%
      \csname LTb\endcsname%
      \put(814,4819){\makebox(0,0)[r]{\strut{}$100$}}%
      \csname LTb\endcsname%
      \put(1191,484){\makebox(0,0){\strut{}$20$}}%
      \csname LTb\endcsname%
      \put(1805,484){\makebox(0,0){\strut{}$40$}}%
      \csname LTb\endcsname%
      \put(2418,484){\makebox(0,0){\strut{}$60$}}%
      \csname LTb\endcsname%
      \put(3031,484){\makebox(0,0){\strut{}$80$}}%
      \csname LTb\endcsname%
      \put(3645,484){\makebox(0,0){\strut{}$100$}}%
      \csname LTb\endcsname%
      \put(4258,484){\makebox(0,0){\strut{}$120$}}%
      \csname LTb\endcsname%
      \put(4871,484){\makebox(0,0){\strut{}$140$}}%
      \csname LTb\endcsname%
      \put(5484,484){\makebox(0,0){\strut{}$160$}}%
      \csname LTb\endcsname%
      \put(6098,484){\makebox(0,0){\strut{}$180$}}%
      \csname LTb\endcsname%
      \put(6711,484){\makebox(0,0){\strut{}$200$}}%
    }%
    \gplgaddtomacro\gplfronttext{%
      \csname LTb\endcsname%
      \put(209,2761){\rotatebox{-270}{\makebox(0,0){\strut{}$\tau_\text{int}$}}}%
      \put(3874,154){\makebox(0,0){\strut{}$L$}}%
      \csname LTb\endcsname%
      \put(4305,4591){\makebox(0,0)[r]{\strut{}FA, $T=\nicefrac\pi2$}}%
      \csname LTb\endcsname%
      \put(4305,4261){\makebox(0,0)[r]{\strut{}no FA, $T=\num{0.6}$}}%
    }%
    \gplbacktext
    \put(0,0){\includegraphics{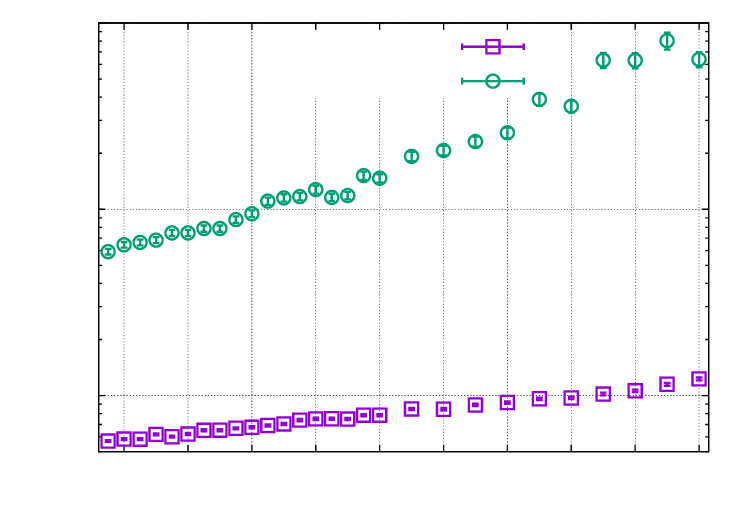}}%
    \gplfronttext
  \end{picture}%
\endgroup
\begingroup
  \inputencoding{latin1}%
  \makeatletter
  \providecommand\color[2][]{%
    \GenericError{(gnuplot) \space\space\space\@spaces}{%
      Package color not loaded in conjunction with
      terminal option `colourtext'%
    }{See the gnuplot documentation for explanation.%
    }{Either use 'blacktext' in gnuplot or load the package
      color.sty in LaTeX.}%
    \renewcommand\color[2][]{}%
  }%
  \providecommand\includegraphics[2][]{%
    \GenericError{(gnuplot) \space\space\space\@spaces}{%
      Package graphicx or graphics not loaded%
    }{See the gnuplot documentation for explanation.%
    }{The gnuplot epslatex terminal needs graphicx.sty or graphics.sty.}%
    \renewcommand\includegraphics[2][]{}%
  }%
  \providecommand\rotatebox[2]{#2}%
  \@ifundefined{ifGPcolor}{%
    \newif\ifGPcolor
    \GPcolortrue
  }{}%
  \@ifundefined{ifGPblacktext}{%
    \newif\ifGPblacktext
    \GPblacktexttrue
  }{}%
  \let\gplgaddtomacro\g@addto@macro
  \gdef\gplbacktext{}%
  \gdef\gplfronttext{}%
  \makeatother
  \ifGPblacktext
    \def\colorrgb#1{}%
    \def\colorgray#1{}%
  \else
    \ifGPcolor
      \def\colorrgb#1{\color[rgb]{#1}}%
      \def\colorgray#1{\color[gray]{#1}}%
      \expandafter\def\csname LTw\endcsname{\color{white}}%
      \expandafter\def\csname LTb\endcsname{\color{black}}%
      \expandafter\def\csname LTa\endcsname{\color{black}}%
      \expandafter\def\csname LT0\endcsname{\color[rgb]{1,0,0}}%
      \expandafter\def\csname LT1\endcsname{\color[rgb]{0,1,0}}%
      \expandafter\def\csname LT2\endcsname{\color[rgb]{0,0,1}}%
      \expandafter\def\csname LT3\endcsname{\color[rgb]{1,0,1}}%
      \expandafter\def\csname LT4\endcsname{\color[rgb]{0,1,1}}%
      \expandafter\def\csname LT5\endcsname{\color[rgb]{1,1,0}}%
      \expandafter\def\csname LT6\endcsname{\color[rgb]{0,0,0}}%
      \expandafter\def\csname LT7\endcsname{\color[rgb]{1,0.3,0}}%
      \expandafter\def\csname LT8\endcsname{\color[rgb]{0.5,0.5,0.5}}%
    \else
      \def\colorrgb#1{\color{black}}%
      \def\colorgray#1{\color[gray]{#1}}%
      \expandafter\def\csname LTw\endcsname{\color{white}}%
      \expandafter\def\csname LTb\endcsname{\color{black}}%
      \expandafter\def\csname LTa\endcsname{\color{black}}%
      \expandafter\def\csname LT0\endcsname{\color{black}}%
      \expandafter\def\csname LT1\endcsname{\color{black}}%
      \expandafter\def\csname LT2\endcsname{\color{black}}%
      \expandafter\def\csname LT3\endcsname{\color{black}}%
      \expandafter\def\csname LT4\endcsname{\color{black}}%
      \expandafter\def\csname LT5\endcsname{\color{black}}%
      \expandafter\def\csname LT6\endcsname{\color{black}}%
      \expandafter\def\csname LT7\endcsname{\color{black}}%
      \expandafter\def\csname LT8\endcsname{\color{black}}%
    \fi
  \fi
    \setlength{\unitlength}{0.0500bp}%
    \ifx\gptboxheight\undefined%
      \newlength{\gptboxheight}%
      \newlength{\gptboxwidth}%
      \newsavebox{\gptboxtext}%
    \fi%
    \setlength{\fboxrule}{0.5pt}%
    \setlength{\fboxsep}{1pt}%
\begin{picture}(7200.00,5040.00)%
    \gplgaddtomacro\gplbacktext{%
      \csname LTb\endcsname%
      \put(682,1477){\makebox(0,0)[r]{\strut{}$1$}}%
      \csname LTb\endcsname%
      \put(682,4046){\makebox(0,0)[r]{\strut{}$10$}}%
      \csname LTb\endcsname%
      \put(2421,484){\makebox(0,0){\strut{}$3$}}%
      \csname LTb\endcsname%
      \put(3435,484){\makebox(0,0){\strut{}$6$}}%
      \csname LTb\endcsname%
      \put(5789,484){\makebox(0,0){\strut{}$30$}}%
      \csname LTb\endcsname%
      \put(814,484){\makebox(0,0){\strut{}$1$}}%
      \csname LTb\endcsname%
      \put(4182,484){\makebox(0,0){\strut{}$10$}}%
    }%
    \gplgaddtomacro\gplfronttext{%
      \csname LTb\endcsname%
      \put(209,2761){\rotatebox{-270}{\makebox(0,0){\strut{}$\tau_\text{int}$}}}%
      \put(3808,154){\makebox(0,0){\strut{}$\beta$}}%
      \csname LTb\endcsname%
      \put(2926,4591){\makebox(0,0)[r]{\strut{}FA, $T=\nicefrac\pi2$}}%
      \csname LTb\endcsname%
      \put(2926,4261){\makebox(0,0)[r]{\strut{}no FA, $T=\nicefrac{1}{\sqrt{\beta}}$  }}%
      \csname LTb\endcsname%
      \put(2926,3931){\makebox(0,0)[r]{\strut{}no FA, $T=1$}}%
    }%
    \gplbacktext
    \put(0,0){\includegraphics{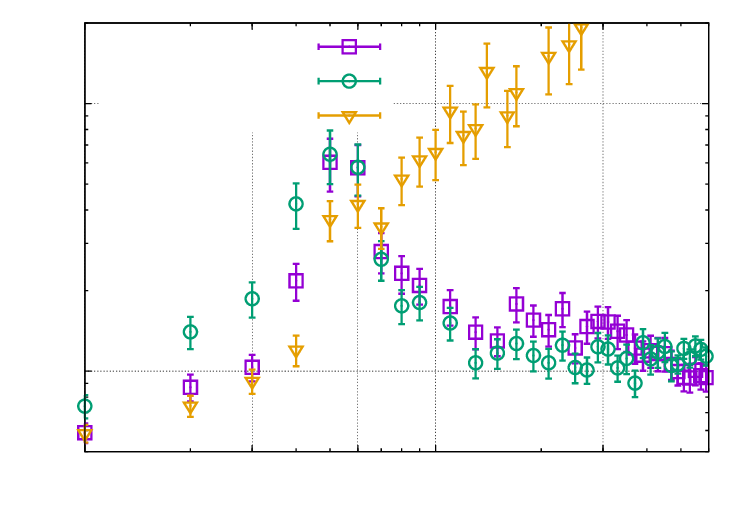}}%
    \gplfronttext
  \end{picture}%
\endgroup
}}
	\caption{Integrated autocorrelation time $\tau_\text{int}$ of the plaquette expectation value $\erwartung{P(\beta)}$ in pure gauge theory HMC simulations. Left: 2D, U$(1)$ weak coupling $\beta=\num{10}$ and different lattice sizes $L$; Right: 4D, $L=10$ lattice, SU$(3)$ and different coupling strengths $\beta$. All simulations required the same compute time per trajectory and volume. Measurements were performed every trajectory. Trajectory lengths without FA were chosen as follows. Left: $T=\num{0.6}$ tuned (by hand) to minimise $\tau_\text{int}$ on the $15\times15$ lattice; Right: $T=1$ because it is the `canonical' choice and $T=\nicefrac{1}{\sqrt{\beta}}$ because of its correct scaling~\cite{ostmeyer2024minimal}.}
	\label{fig:pure_gauge_tau}
\end{figure*}

Lattice gauge theories~\cite{Wilson:1974sk,Gattringer:2010zz} typically encountered in physics are formulated on compact groups are therefore intrinsically highly anharmonic. Apart from the regime of very weak coupling (large $\beta$), the HMC with EFA is therefore not expected to be very efficient. A gauge-invariant version of the HMC with FA has been proposed in Ref.~\cite{DUANE1988101}. An alternative is to expand the pure gauge action to quadratic order in the fields and apply classical FA~\cite{ostmeyer2024minimal,PhysRevD.32.2736}. The latter approach guarantees optimal sampling in the weak coupling limit, but beyond this limit numerical tests (see fig.~\ref{fig:pure_gauge_tau}) remain inconclusive.

\section{Summary}

``The Physicist's Guide to the HMC'' contains a hands-on instruction manual for the correct and efficient use of the HMC~\cite{Duane1987}. For near-normal probability distributions exact Fourier acceleration (EFA)~\cite{ostmeyer2024minimal} leads to optimal sampling as described in \cref{sec:harmonic} and summarised in \cref{alg:efa,alg:efa-hmc,alg:leap-frog}. The HMC is applicable to any probability distribution of continuous variables and the general case is discussed in \cref{sec:anharmonic}. Efficient sampling using the HMC is guaranteed as long as the trajectory length is randomised and chosen long enough on average~\cite{apers2022hamiltonian}. In addition, on non-compact spaces the HMC needs to be combined with radial updates~\cite{original_radial_update,Ostmeyer:2024gnh} as described in \cref{alg:poly_pot}. 
\section*{Code and Data}
All the codes used for this work have been published under open access, see Ref.~\cite{ostmeyer2024minimal}. The analysis used the light-weight tool \texttt{comp-avg}~\cite{comp-avg}. Most of the simulations in this work can be reproduced very quickly, nonetheless the resulting data will gladly be provided upon request.

\section*{Acknowledgements}
The authors thanks Evan Berkowitz, Benjamin Cohen-Stead, Sander Gribling, Anthony Kennedy, Stefan Krieg, Tom Luu, Marcel Rodekamp, and Carsten Urbach for their helpful comments.
Special thanks go to Pavel Buividovich for insightful discussions and his support.
This work was funded in part by the STFC Consolidated Grant ST/T000988/1 and by the Deutsche Forschungsgemeinschaft (DFG,
German Research Foundation) as part of the CRC 1639 NuMeriQS -- project no.\ 511713970.
Numerical simulations were undertaken on Barkla (though simulations with EFA could have easily been run on a laptop), part of the High Performance Computing facilities at the University of Liverpool, UK. 

\FloatBarrier
\bibliographystyle{JHEP}
\bibliography{bibliography}

\providecommand{\href}[2]{#2}\begingroup\raggedright\begin{thebibliography}{10}

\bibitem{Duane1987}
S.~Duane, A.D.~Kennedy, B.J.~Pendleton and D.~Roweth, \emph{{Hybrid Monte
  Carlo}},
  \href{https://doi.org/10.1016/0370-2693(87)91197-X}{\emph{Phys.~Lett.~B}
  {\bfseries 195} (1987) 216 }.

\bibitem{ostmeyer2024minimal}
J.~Ostmeyer and P.~Buividovich, \emph{{Minimal Autocorrelation in Hybrid Monte
  Carlo simulations using Exact Fourier Acceleration}},
  \href{https://arxiv.org/abs/2404.09723}{{\ttfamily 2404.09723}}.

\bibitem{OMELYAN2003272}
I.~Omelyan, I.~Mryglod and R.~Folk, \emph{{Symplectic analytically integrable
  decomposition algorithms: classification, derivation, and application to
  molecular dynamics, quantum and celestial mechanics simulations}},
  \href{https://doi.org/https://doi.org/10.1016/S0010-4655(02)00754-3}{\emph{Computer
  Physics Communications} {\bfseries 151} (2003) 272}.

\bibitem{trotter_omelyan}
J.~Ostmeyer, \emph{{Optimised Trotter decompositions for classical and quantum
  computing}}, \href{https://doi.org/10.1088/1751-8121/acde7a}{\emph{J. Phys.
  A} {\bfseries 56} (2023) 285303}
  [\href{https://arxiv.org/abs/2211.02691}{{\ttfamily 2211.02691}}].

\bibitem{PhysRevD.32.2736}
G.G.~Batrouni, G.R.~Katz, A.S.~Kronfeld, G.P.~Lepage, B.~Svetitsky and
  K.G.~Wilson, \emph{Langevin simulations of lattice field theories},
  \href{https://doi.org/10.1103/PhysRevD.32.2736}{\emph{Phys. Rev. D}
  {\bfseries 32} (1985) 2736}.

\bibitem{Takaishi:1999bi}
T.~Takaishi, \emph{{Choice of integrator in the hybrid Monte Carlo algorithm}},
  \href{https://doi.org/10.1016/S0010-4655(00)00161-2}{\emph{Comput. Phys.
  Commun.} {\bfseries 133} (2000) 6}
  [\href{https://arxiv.org/abs/hep-lat/9909134}{{\ttfamily hep-lat/9909134}}].

\bibitem{Neal:2011mrf}
R.M.~Neal, \emph{{Handbook of Markov Chain Monte Carlo}}, {Chapman and
  Hall/CRC} (5, 2011), \href{https://doi.org/10.1201/b10905}{10.1201/b10905},
  [\href{https://arxiv.org/abs/1206.1901}{{\ttfamily 1206.1901}}].

\bibitem{Ostmeyer:2023azi}
J.~Ostmeyer, T.~Nematiaram, A.~Troisi and P.~Buividovich,
  \emph{{First-principles quantum Monte Carlo study of charge-carrier mobility
  in organic molecular semiconductors}},
  \href{https://doi.org/10.1103/PhysRevApplied.22.L031004}{\emph{Phys. Rev.
  Applied} {\bfseries 22} (2024) L031004}
  [\href{https://arxiv.org/abs/2312.14914}{{\ttfamily 2312.14914}}].

\bibitem{WOLFF2004143}
U.~Wolff, \emph{{Monte Carlo errors with less errors}},
  \href{https://doi.org/https://doi.org/10.1016/S0010-4655(03)00467-3}{\emph{Computer
  Physics Communications} {\bfseries 156} (2004) 143}.

\bibitem{Mackenzie:1989us}
P.B.~Mackenzie, \emph{{An Improved Hybrid Monte Carlo Method}},
  \href{https://doi.org/10.1016/0370-2693(89)91212-4}{\emph{Phys. Lett. B}
  {\bfseries 226} (1989) 369}.

\bibitem{apers2022hamiltonian}
S.~Apers, S.~Gribling and D.~Szilágyi, \emph{{Hamiltonian Monte Carlo for
  efficient Gaussian sampling: long and random steps}},  2022.

\bibitem{NUTS-2014}
M.D.~Homan and A.~Gelman, \emph{{The No-U-turn sampler: adaptively setting path
  lengths in Hamiltonian Monte Carlo}}, {\emph{J. Mach. Learn. Res.} {\bfseries
  15} (2014) 1593} [\href{https://arxiv.org/abs/1111.4246}{{\ttfamily
  1111.4246}}].

\bibitem{ising}
J.~Ostmeyer, E.~Berkowitz, T.~Luu, M.~Petschlies and F.~Pittler, \emph{{The
  Ising Model with Hybrid Monte Carlo}},
  \href{https://doi.org/10.1016/j.cpc.2021.107978}{\emph{Comput. Phys. Commun.}
  {\bfseries 265} (2021) 107978}.

\bibitem{onsager_2d_solution}
L.~Onsager, \emph{{Crystal Statistics. I. A Two-Dimensional Model with an
  Order-Disorder Transition}},
  \href{https://doi.org/10.1103/PhysRev.65.117}{\emph{Phys. Rev.} {\bfseries
  65} (1944) 117}.

\bibitem{original_radial_update}
A.~Kennedy and X.~Yu, \emph{{The convergence of HMC on non-compact Riemannian
  Manifolds}},  forthcoming.

\bibitem{Ostmeyer:2024gnh}
J.~Ostmeyer, \emph{{Exponential speed up in Monte Carlo sampling through Radial
  Updates}},  \href{https://arxiv.org/abs/2411.18218}{{\ttfamily 2411.18218}}.

\bibitem{Temmen:2024pcm}
F.~Temmen, E.~Berkowitz, A.~Kennedy, T.~Luu, J.~Ostmeyer and X.~Yu,
  \emph{{Overcoming Ergodicity Problems of the Hybrid Monte Carlo Method using
  Radial Updates}},  in \emph{{41st International Symposium on Lattice Field
  Theory}}, 10, 2024 [\href{https://arxiv.org/abs/2410.19148}{{\ttfamily
  2410.19148}}].

\bibitem{Temmen_ergodicity}
F.~Temmen, E.~Berkowitz, A.~Kennedy, T.~Luu, J.~Ostmeyer and X.~Yu,
  \emph{{Fully ergodic simulations of the Hubbard model using Radial Updates}},
   forthcoming.

\bibitem{Wilson:1974sk}
K.G.~Wilson, \emph{{Confinement of Quarks}},
  \href{https://doi.org/10.1103/PhysRevD.10.2445}{\emph{Phys. Rev. D}
  {\bfseries 10} (1974) 2445}.

\bibitem{Gattringer:2010zz}
C.~Gattringer and C.B.~Lang, \emph{{Quantum chromodynamics on the lattice}},
  vol.~788, Springer, Berlin (2010),
  \href{https://doi.org/10.1007/978-3-642-01850-3}{10.1007/978-3-642-01850-3}.

\bibitem{DUANE1988101}
S.~Duane and B.J.~Pendleton, \emph{Gauge invariant fourier acceleration},
  \href{https://doi.org/https://doi.org/10.1016/0370-2693(88)91270-1}{\emph{Physics
  Letters B} {\bfseries 206} (1988) 101}.

\bibitem{comp-avg}
J.~Ostmeyer, \emph{{comp-avg: Compare Averages of time series and more}},
  \href{https://zenodo.org/records/13479395}{10.5281/zenodo.10794620 (2024)}.

\end{thebibliography}\endgroup

\end{document}